\documentclass[12pt,preprint]{aastex}

\begin{document}

\title{A Unified Fitting of \ion{H}{1} and \ion{He}{2} Ly${\alpha}$
Transmitted Flux of QSO HE2347 with $\Lambda$CDM Hydrodynamic
Simulations}

\author{Jiren Liu\altaffilmark{1},
   Priya Jamkhedkar\altaffilmark{2},
     Wei Zheng\altaffilmark{3},
Long-Long Feng\altaffilmark{4,5},
       and Li-Zhi Fang\altaffilmark{2}}

\altaffiltext{1}{Center for Astrophysics, University of Science
and Technology of China, Hefei, Anhui 230026,P.R.China}
\altaffiltext{2}{Department of Physics, University of Arizona,
Tucson, AZ 85721}
\altaffiltext{3}{Department of Physics and Astronomy, Johns
Hopkins University, MD 21218}
\altaffiltext{4}{Purple Mountain Observatory, Nanjing, 210008,
P.R. China.}
\altaffiltext{5}{National
Astronomical Observatories, Chinese Academy of Science, Chao-Yang
District, Beijing 100012, P.R. China}

\begin{abstract}

Using cosmological hydrodynamic simulations of the LCDM model, we
present a  comparison between the simulation sample and real data
sample of \ion{H}{1} and \ion{He}{2} Ly$\alpha$ transmitted flux
in the absorption spectra of the QSO HE2347-4342. The $\Lambda$CDM
model is successful in simultaneously explaining the
statistical features of both \ion{H}{1} and \ion{He}{2} Ly$\alpha$
transmitted flux.  It includes: 1.) the power spectra
of the transmitted flux of \ion{H}{1} and \ion{He}{2}
can be well fitted on all scales $\geq 0.28$ h$^{-1}$ Mpc for H,
and $\geq 1.1$ h$^{-1}$ Mpc for He; 2.) the Doppler parameters of
absorption features of \ion{He}{2} and \ion{H}{1} are found to be
turbulent-broadening; 3.) the ratio of \ion{He}{2} to \ion{H}{1}
optical depths are substantially scattered, due to the significant
effect of noise.
A large part of the $\eta$-scatter  is due to the noise in
the \ion{He}{2} flux.
However, the real data contain more low-$\eta$ events than
simulation sample. This discrepancy may indicate that the
mechanism leading extra fluctuations upon the simulation data,
such as a fluctuating UV radiation background, is needed. Yet,
models of these extra fluctuations should satisfy the constraints:
1.) if the fluctuations are Gaussian,  they should be limited by the
power spectra of observed \ion{H}{1} and \ion{He}{2} flux;
2.) if the fluctuations are non-Gaussian, they should be limited by
the observed non-Gaussian features of the \ion{H}{1} and \ion{He}{2} flux.

\end{abstract}


\section{Introduction}

Ly$\alpha$ forest lines and transmitted flux of QSOs' absorption
spectra provide the most valuable samples in studying the physical
state of the intergalactic medium (IGM) and gravitational
clustering at high redshifts. High resolution samples of QSOs'
Ly$\alpha$ absorption spectra are important to test models of
cosmic structure formation on small scales. Recently, the
\ion{H}{1} Ly$\alpha$ transmitted flux of HE2347-4342 has been
used to compare with hydrodynamical simulation samples of the
$\Lambda$CDM model \citep[][here after Paper I]{jam05}. The
results suggest that the $\Lambda$CDM model is successful in
explaining the power spectrum and intermittency of the HE2347-4342
sample. There is no discrepancy between the simulated and observed
flux fields with regards to their statistical behavior from the
second to the eighth orders and till the comoving scales as small
as about 0.28 h$^{-1}$ Mpc. This result seems not to support the
necessity of reducing the power of density perturbations relative
to the standard $\Lambda$CDM model on small scales, as was implied
by the lack of dense cores in the halo's center given by the
so-called universal density profile
\citep{fp94,sw03,mc03,zb03,sim03}

The IGM is also traced by \ion{He}{2} Ly$\alpha$ absorption. Because the
ionizing threshold of \ion{He}{2} is high (54.4 eV), and recombination rate
of \ion{He}{3} is also high, the \ion{He}{2} Ly$\alpha$ absorption of IGM
generally is much stronger than \ion{H}{1} Ly$\alpha$. Therefore, it is
expected that the \ion{He}{2} Ly$\alpha$ forest and transmitted flux of
high redshift QSOs can play a similar role as \ion{H}{1} forests in
constraining cosmological models, and can even yield stronger
constraints on the models than \ion{H}{1} forests \citep{zhang95,croft97}.
However, due to the lack of \ion{He}{2} data,
the comparison of \ion{He}{2} spectra between
model predictions and observations could not be made in a similar way as
for \ion{H}{1} Ly$\alpha$ absorption spectra.
Thanks to the $FUSE$ data of HE2347-4342, we can obtain
moderate resolution spectra of \ion{He}{2} Ly$\alpha$ transmitted flux
in the redshift range $2.0<z<2.9$ \citep{k01}.
It provides the possibility of making a similar analysis
as for the \ion{H}{1} transmitted flux.

We cannot simply repeat the analysis as Paper I, because of
a new problem: the ratio between the optical
depths of \ion{He}{2} and \ion{H}{1}, $\eta = 4\tau_{\rm He
II}/\tau_{\rm HI}$. If assuming 1.) the effect of thermal
broadening and peculiar velocities of IGM are negligible,
and 2.) \ion{He}{2} and \ion{H}{1} are in photoionization equilibrium,
the ratio $\eta$ should basically be constant,
and have a low level of scatter because $\eta$ is
weakly dependent on the temperature of IGM.
Observations reveal, however, that $\eta$ is significantly
scattered from pixel to pixel, or from line to line
\citep{k01,sm02}. For the $FUSE$ data of HE2347-4342, the scatter
of $\eta$ is from 1 to a few hundreds \citep{sh04}, and even as
high as $\simeq 10^4$ \citep{zh04}.

The large $\eta$-scatter has been used as an indicator of the
inhomogeneity of UV radiation background caused by the radiation
transfer (RT) of the UV photons in a nonuniform density field. A
3-dimensional radiation transfer calculation on the shadowing,
self-shielding and filtering predicted that the mean of $\eta$
should be as large as $\langle \eta \rangle > 200$ \citep{mf05}.
However, the observed result is $\langle \eta \rangle < 100$
\citep{zh04,sh04}. Therefore, the RT effect explanation of $\eta$
scatter is far from settling. This result motivated us to
reconsider the assumptions above mentioned. If above-mentioned
assumptions 1.) and 2.) are not hold, the $\eta$-scatter would not
be a direct measurement of the inhomogeneity of UV background.

In this paper, we will study the effect of thermal broadening
and peculiar velocities on the \ion{H}{1} and \ion{He}{2}
Ly${\alpha}$ transmitted flux. It has been shown recently that in
nonlinear regime the velocity field $v(x)$ of cosmic baryon matter
consists of strong shocks on various spatial scales and in high
and low mass density area \citep{Kim05}. The statistical behavior
of the velocity field is similar to a fully developed turbulence
\citep{he06}. Therefore, the effect of thermal broadening
and peculiar velocities of IGM would not be negligible. We should, at
least, estimate the imprints of the non-trivial velocity field on
the Ly$\alpha$ forests of \ion{He}{2} and \ion{H}{1}.

The paper is organized as follows. Section 2 describes the
observed data of HE2347-4342. Section 3 presents the method
to simulate the Ly$\alpha$ forests of \ion{He}{2} and \ion{H}{1}.
As in Paper I, we use the WIGEON method of cosmological
hydrodynamic simulations to produce the simulation samples,
as this code is especially effective in
capturing singular and complex structures \citep{feng04}. On the
other hand, \ion{He}{2} would be formed in hotter gas, such as
shock heated IGM \citep{cen99,dave01}, a method of effectively
simulating shocks would be important. Velocity field effect on
absorption width is shown in Section 4. Section 5 presents the
analysis of the ratio between the optical depths of \ion{He}{2}
and \ion{H}{1} Ly$\alpha$ transmissions. The power spectrum
of the \ion{H}{1} and \ion{He}{2} flux will be
discussed in \S 6. Discussions and conclusions are
given in \S 7.

\section{Data of HE2347-4342}

The data of \ion{H}{1} Ly$\alpha$ transmitted flux used in this paper is
the same as in Paper I. The $FUSE$ data of the \ion{He}{2} Ly$\alpha$
transmitted flux of HE2347-4342 is described in \citet{zh04}.
The wavelength region is $904 - 1188$ \AA, which corresponds
to the redshift range $2.0 \leq z \leq 2.9$, as the wavelength
303.78 \AA\ of \ion{He}{2} Ly$\alpha$ in rest frame.
The spectrum has a constant bin size of $\Delta \lambda = 0.025$~\AA.
In terms of the local velocity, the resolution is $dv
\simeq 8.3 - 6.3$ km s$^{-1}$, and mean $d v \simeq 7$
km~s$^{-1}$. The mean $S/N$ is 2.14. Following the approach of
\citet{sh04}, we bin the data into
$\Delta \lambda = 0.05$~\AA~to reduce uncertainties
in the effective spectrograph resolution and oversampling effect.
The distance between $N$ pixels in the units of the local velocity
scale is given by $\delta v=2c[1-\exp(-N d v/2c)]$~km~s$^{-1}$,
corresponding to comoving scale $D = \delta v (1+z)/H(z)$.

The flux in 2729 pixels, i.e., 24\% of the total pixels, are less
than zero. Obviously, the points with negative flux is unphysical,
it should be excluded in the statistics below. For sample deleting
all the pixels with negative flux, the mean transmission about
0.4, or effective optical depth $\sim 0.9$. The optical depths of
\ion{He}{2}
over the ranges from 0.1 to 2.3 are with 10\% uncertainties. A
better statistical measurement of the fluctuations of flux is
given by the ratio between the optical depths of \ion{He}{2}, $\tau_{\rm
HeII}$, and \ion{H}{1}, $\tau_{\rm HI}$, for each pixel. The distribution
of $\eta$ is scattered in the range from 0.1 to about 500, while
the mean of $\eta$ is $\simeq 80$ \citep{sh04}.

We also assume that all absorption in the $FUSE$ spectrum is due
to \ion{He}{2}, although it is subject to metal-line contaminations.
Generally identified metal-lines are connected with a Lyman-limit
system \citep{sm02}. The Doppler width of metal lines are
generally narrow with $\delta v \leq~ 20~{\rm km~s^{-1}}$. In this
paper, we restrict our analysis only to scales $\delta v \geq
30~{\rm km~s^{-1}}$ where metal-line contaminations
is low \citep{hu95,bok03,kim04}.

\section{Hydrodynamic simulation sample}

\subsection{Method}

We use the cosmological hydrodynamic simulation samples produced by
the Weno for Intergalactic medium and Galaxy Evolution and formatiON
(WIGEON) code developed by \citet{feng04}. It is a hybrid
hydrodynamic/$N$-body simulation, consisting of the WENO algorithm
\citep{js96} for baryonic fluid, and $N$-body simulation for
particles of dark matter. The baryon fluid obeys the Navier-Stokes
equation, and is gravitationally coupled with collisionless dark
matter. We have assumed a standard $\Lambda$CDM model, which is
specified by the matter density parameter $\Omega_{\rm m}=0.27$,
baryonic matter density parameter $\Omega_{\rm b}=0.044$,
cosmological constant $\Omega_{\Lambda}=0.73$, Hubble constant
$h=0.71$, the mass fluctuation $\sigma_8=0.84$, and scale-free
spectrum index $n=1$. The ratio of specific heats of the IGM is
$\gamma=5/3$.  The transfer function is calculated using CMBFAST
\citep{sz96}.

The simulation was performed in a periodic, cubic box of size 50
$h^{-1}$Mpc with a 512$^3$ grid and an equal number of dark matter
particles. It starts at redshift $z=99$. A uniform UV-background
of ionizing photons is switched on at $z = 6$ to heat the gas and
reionize the universe. To mimic the enhancement of temperature
due to radiation transfer effects \citep{abel99},
a thermal energy of gas with $T=2\times 10^4$
K is added in the total energy at $z=6$. The clumpy universe would
reprocess the photon spectrum from ionizing sources. The
reprocessing UV spectrum has been calculated by \citet{hm96}. We
use an ionizing background model including QSOs and galaxies with
10\% ionizing photons escape fraction (kindly provided by F.
Haardt). At $z = 2.5$, such an ionizing background produces the
transmission flux of \ion{H}{1} and \ion{He}{2} similar to
observation and an average $\eta$ $\simeq 72$, which is very close
to observed value.

The atomic processes in the plasma of hydrogen and helium of
primordial composition, including ionization, radiative cooling
and heating, and the fraction of \ion{H}{1} and \ion{He}{2} are
calculated in the same way as \citet{th98}. That is, under the
``optically thin'' approximation, once density and temperature of
baryon gas are given, the ionizing state of H and He is directly
determined from the ionization-equilibrium equation.

\subsection{Samples of Ly$\alpha$ transmitted flux}

For given sample of the fields of density, temperature and
velocity of the baryon matter, the optical depth of \ion{H}{1} or
\ion{He}{2} can be produced by a convolution with Voigt profile as
follows \citep{bi95, zhang97}
\begin{equation}
\tau_i(z)=\sigma_i c \int dx
n_{i}(x)\frac{1}{\sqrt{\pi}H(z)b^T_i}V[\frac{\delta z}{b^T_i(1+z)}+
\frac{v(x)}{b^T_i},
b^T_i]
\end{equation}
where $i=$\ion{H}{1} or \ion{He}{2}, $\sigma_i$ is the absorption
cross section of Ly$\alpha$ line, $n_{i}(x)$ the number density,
$\delta z$ is the redshift difference between $z$ and $x$, $v(x)$
the peculiar velocity in unit $c$ and $b^T_i=(2kT/m_ic^2)^{1/2}$
the thermal velocity. $V$ is the Voigt profile, which is
normalized $\int dx (1/\sqrt{\pi}b)V[\frac{\delta
z}{b(1+z)}+\frac{v(x)}{b}, b]=1$. The Hubble constant at redshift
$z$ is $H(z)=H_0\sqrt{\Omega_m(1+z)^3+\Omega_{\Lambda}}$. Eq.(1)
shows, when the terms of thermal broadening and peculiar velocity
are not negligible, the Ly$\alpha$ transmission flux depends on
mass density field $n_{\rm HI}$, $n_{\rm HeII}$ {\it as well as}
the fields of temperature and velocity of IGM.

We produced 100 mock samples of \ion{H}{1} and \ion{He}{2}
Ly$\alpha$ transmitted flux at $z=2.5$ with randomly selected
lines of sight. Each mock spectrum is sampled using $2^{10}$
pixels with the same spectral resolution as the observation. As
the corresponding comoving scale for $2^{10}$ pixels is larger
than the simulation box size, we replicate the sample
periodically. We add Gaussian noise to \ion{H}{1} sample with
signal-to-noise ratio, S/N=50, while \ion{He}{2} with S/N=3.
Figure 1 shows typical samples of \ion{H}{1} and \ion{He}{2}
Ly$\alpha$ transmitted flux fields.

\section{Line width of \ion{He}{2} and \ion{H}{1} Ly$\alpha$
absorptions}

To demonstrate the importance of the velocity field, we consider
the line widths of Ly$\alpha$ absorption. If the Ly$\alpha$
absorption lines are purely thermal broadened, the line width, or
Doppler parameter, $b$(\ion{He}{2}) of \ion{He}{2} should be less
than $b$(\ion{H}{1}) by a factor of 2. However, \citet{zh04} found
$b$(\ion {He}{2})$=\xi b$(\ion{H}{1}) and $\xi \simeq 1$. They
concluded that the velocity field in IGM is dominated by
turbulence.

This result can be explained with considering the velocity field
in eq.(1). A flux field given by eq.(1) generally is not given by
a superposition of lines with Gaussian profile, because the
peculiar velocity $v(x)$ is a random field. Therefore, the line
width given by Gaussian profile fitting is not always equal to the
thermal broadening $b^T$. For instance, when $b^T_i$ is small, the
Voigt profile will picks up only pixels with $\delta z +
(1+z)v(x)\simeq 0$. In this case, the line width can be estimated
by $(1+z)\langle(\Delta v)^2\rangle^{1/2}$, i.e. the line width is
dominated by the variance of the velocity field, which is the same
for \ion{He}{2} and \ion{H}{1}. This is turbulent broadening. It
has been shown recently that the velocity field $v(x)$ shows the
feature of a fully developed turbulence \citep{he06}.

To test this point, we identify the absorption lines by the
similar way as that for the real sample. We use AUTOVP code
\citep{dave97} to decompose the transmitted flux with Gaussian
profile, and estimate the parameters of line width, column
density, and the centroid wavelength of each line. Figure 2 shows
$b$(\ion{H}{1}) vs. $b$(\ion{He}{2}) for both real data and
simulation samples. The real data are taken from \citet{zh04}.
Obviously, the plot of $b$(\ion{H}{1}) vs.
$b$(\ion{He}{2}) from simulated data does not follow the thermal
broadening relation $b$(\ion{H}{1}) = 2$b$(\ion{He}{2}), but
similar to  the turbulent broadening.

\section{The ratio between \ion{He}{2} and \ion{H}{1} Ly$\alpha$
optical depths}

\subsection{The scatter of optical depth ratio}

If the Voigt profile can be approximated by a Dirac delta
function, eq.(1) yields
\begin{equation}
\frac{\tau_{\rm HeII}(z)}{\tau_{\rm HI}(z)}=\frac{1}{4}
\frac{n_{\rm HeII}(z)}{n_{\rm HI}(z)},
\end{equation}
If atoms and ions of \ion{H}{1}, \ion{H}{2}, \ion{He}{1}, \ion{He}{2}
and
\ion{He}{3} are in state of photoionization equilibrium, the ratio of
\ion{He}{2} to \ion{H}{1} is \citep{fardal98}
\begin{equation}
\frac{n_{\rm HeII}(z)}{n_{\rm HI}(z)}\simeq 1.70
 \frac{J_{\rm HI}}{J_{\rm HeII}}\frac{3+\alpha_4}
 {3+\alpha_1}\left (\frac{T(z)}{10^{4.3}}\right )^{0.06} ,
\end{equation}
where we assume that the UV radiation backgrounds around
wavelengths wavelength $c/\nu_{\rm HI}=912$\AA~ and $c/\nu_{\rm
HeII}=228$ \AA~ are, respectively, $J_{\nu}=J_{\rm
HI}(\nu/\nu_{\rm HI})^{-\alpha_1}$ and $J_{\nu}=J_{\rm
HeII}(\nu/\nu_{\rm HeII})^{-\alpha_4}$, the parameters $J_{\rm
HI}$ and $J_{\rm HeII}$ being the specific intensities, $\alpha_1$
and $\alpha_4$ the index of their power laws.

If the UV radiation background is spatially uniform, i.e. the
parameters $J_{\rm HI}$, $J_{\rm HeII}$, ${\alpha_1}$ and
${\alpha_4}$ are constant, the ratio $n_{\rm HeII}(z)/n_{\rm
HI}(z)$ of eq.(3) is approximately spatially constant, because the
temperature-dependence ($T^{0.06}$) is very weak. Thus, from
eqs.(2) and (3) we may expect that the ratio of optical depths
$\tau_{\rm HeII}(z)/\tau_{\rm HI}(z)$ should basically be
constant, i.e. the scatter of ratio $\tau_{\rm HeII}(z)/\tau_{\rm
HI}(z)$ with respect to its mean $\langle \tau_{\rm
HeII}(z)/\tau_{\rm HI}(z)\rangle$ has to be very small.

Thus, the observed $\eta$ scatter \citep{k01,sm02,zh04,sh04} may
indicate that the ratios, $J_{\rm HI}/J_{\rm HeII}$ and
$(3+\alpha_4)/(3+\alpha_1)$ are not constant, but significantly
different from pixel to pixel, or from line to line. However, eq.(2)
is based on the assumption that the effects of thermal broadening and
peculiar velocities are ignored. As has been shown in last section, the
effects of thermal broadening and peculiar velocities may not be
always small. Therefore, we should estimate the $\eta$-scatter caused
by thermal broadening and peculiar velocities.

Since the assumption of eq.(2) is not hold,
we use $\eta$ to stand only for $4\tau_{\rm
HeII}/\tau_{\rm HI}(z)$ below. Figure 3 illustrates the effect of
thermal broadening and a peculiar velocity field. It shows 1.) top
panel: the 1-D distributions of the ratio $\eta$ of optical depths
given by eq.(1), 2.) second panel: $\eta$ distribution from optical
depths of eq.(1), but using a delta function for the Voigt profile; 3.)
third panel: $\eta$ given by optical depths of eq.(1) and adding,
respectively, S/N=3 and S/N=50 noises to the \ion{He}{2} and \ion{H}{1}
transmitted flux; 4.) forth panel: temperature $T$ distribution; 5.)
bottom panel: mass density field of baryon matter.

From the panel 2 of Fig. 3 one can see clearly that if the effect
of thermal broadening is ignored, i.e. the Voigt profile is
approximated by a Dirac delta function, $\eta$ is almost a
constant in the whole range. In this approximation, the ratio
$\eta$ does not depend on the fluctuations of temperature, mass
density and velocity of baryon gas, but only on the density ratio
$n_{\rm HII}/n_{\rm HI}$. The ratio $n_{\rm HII}/n_{\rm HI}$, however,
always keeps constant, even when $n_{\rm HII}$ and $n_{\rm HI}$
fluctuate strongly. The $\eta$ distribution shown in panel 2 contains
a significant sharp decline at the highest temperature region. It
probably results from dielectronic recombination of \ion{He}{2} that
is dominant at such high temperatures, and reduces the number
$n_{\rm HeII}$. At less high temperatures, the collision ionization
rate of \ion{H}{1} is higher than \ion{He}{2}, accordingly, there
appear some small bumps as those visualized in panel 2.

However, the top panel shows that, when thermal broadening is
included, the fluctuations of $\eta$ are higher at the positions
with higher temperature and higher density. These fluctuations are
due partially to the difference between the thermal velocities of
\ion{He}{2} and \ion{H}{1}. It leads to the distribution of
\ion{H}{1} to be more extended than \ion{He}{2}. The
hydrodynamical velocity field $v(x)$ is the same for \ion{He}{2}
and \ion{H}{1}. It doesn't cause the change of the ratio $n_{\rm
HII}/n_{\rm HI}$. The effect of thermal broadening on $\eta$
scatter has also been noted by \citet{croft97}, but in their
samples the scatter caused by thermal broadening is not very
significant. It is probably because more shocks are resolved on
small scales in the WIGEON simulations, which leads to stronger
temperature fluctuations \citep{he04}. The panel 3 of Fig. 3 shows
more significant scatter of $\eta$ if the noise is added. From
Figure 1, we have seen that the transmitted flux of \ion{He}{2} is
substantially affected by the S/N=3 noise.

\subsection{PDF of $\eta$}

We now examine the probability distribution function (PDF) of
$\eta$, which is calculated pixel by pixel as \citet{sh04}. Figure
4 shows the PDFs for 1.) real data; 2.) simulation sample of the
transmitted flux from eq.(1);  3.) simulation samples of the
transmitted flux from eq.(1), and adding, respectively, S/N=3 and
S/N=50 noise to the \ion{He}{2} and \ion{H}{1} transmitted flux,
and 4.) simulated samples without thermal broadening, but adding noise.
The simulation PDFs are calculated with 100 samples of 1-D
transmitted flux of \ion{He}{2} and \ion{H}{1}. The error bars are
the maximum and minimum from 100 independent noise realizations.
The simulation data are also binned into 0.05\AA~to match
observation.

We see that the PDF of the simulated data without adding noise
has the peak at $\log \eta \simeq 1.9$ which is about the same as
real data \citep{zh04,sh04}. However, the width of PDF of the
simulated data without adding noise is much less than the real data.
The maximum scatter of $\eta$ of the simulated data is about
2$\langle \eta \rangle$. The factor 2 is just the
difference between the thermal velocities of \ion{H}{1} and \ion{He}{2}.
Though there are some pixels with  $\eta > 2\langle \eta \rangle$
caused by collision ionization, it is still less than the observed
scatter.

The PDF of simulation data is significantly improved with adding
noise. It looks similar to the observed result. The PDF of
simulation data has the same peak as real data, and the width of
PDF is also about the same as real one. Therefore, the effect of
data noise is substantial for the scatter of $\eta$. This effect
is especially serious on the high $\eta$ events. Samples without
noise contamination do not have events of $\eta > 300$, but the
samples with noise do. This result is about the same as
\citet{sh04}, in which events of $\eta > 460$ are dropped, because
they may largely be from the uncertainty in measuring.

Nevertheless, we see that the PDF of simulation sample with noise
is still lower than the real data in the range $\eta<10$. This
result is also similar to \citet{sh04}. They found that the events
of $\eta<10$ of real data show an factor 2 excess to a Monte Carlo
calculation. Figure 4 also shows that the number of $\eta < 10$ of
real data is about twice as large as simulation sample. However,
\citet{sh04} found that the events of $\eta<30$ of real data are
also more than their Monte Carlo estimation by a factor 2, while
Figure 3 doesn't show so large difference. This is probably
because our simulation contains the effect of thermal broadening
and peculiar velocity field. This point can be seen from the PDF
of $\eta$ for simulation samples without thermal broadening, but
adding noise (the filled dots in Figure 4). It shows there are
more low-$\eta$ events without thermal broadening. In a word, our
simulation result indicates the excess of low-$\eta$ events in
real data, but the difference between real and simulation data is
less than the Monte Carlo estimation.

The effect of noise can also be seen in Figure 5, which presents
the relation between $\eta$ and $\tau_{\rm HI}$ in the range
$\tau_{\rm HI}> 0.01$. The top panel is from real data. The middle
is given by simulation samples with adding noise of S/N=3 to
\ion{He}{2} and S/N=50 to \ion{H}{1}. The bottom shows simulation
samples without noise. The $\eta$-$\tau_{\rm HI}$ distribution of
simulation sample with noise is similar to the real one. The
high-$\eta$ events at low $\tau_{\rm HI}$ area are mainly from
noise. This is consistent with the observed result that $\eta$ is
large in void area $\tau_{\rm HI}< 0.05$ \citep{sh04}.

Figure 6 presents the relation between $\eta$ and column density
$N({\rm HI})$ for real data and simulation samples with noise.
It shows the correlation between $\eta$ and density of \ion{H}{1}:
$\eta$ is larger for lower column density $N({\rm HI})$, and lower
for higher $N({\rm HI})$. This phenomenon is also shown in the
\ion{He}{2} spectrum of HS1700+6416 \citep{rm05}.

\section{Power spectrum}

We now compare the power spectra of H and He transmitted flux of
simulation sample and real data. We calculate the power spectrum
by the same method as Paper I. It gives easy to compare the power
spectra of H and He. Moreover, the real data of He is highly
noisy, for some pixels, the S/N ratio is lower than 1, and some
pixels with negative flux. To eliminate the effect of these pixels
on the power spectrum calculation, we should use the algorithm of
denoising or conditional-counting as Paper I
\citep{donoho95,jam01}.

The power spectra of \ion{H}{1} and \ion{He}{2}  Ly$\alpha$
transmitted flux of HE2347 are shown in Figure 7. We plot the
power spectrum of \ion{H}{1} in top panel as an indicator of the
goodness of the simulation sample of this paper, which is produced
in a box with size larger than that of Paper I by a factor of 4.
The parameters $f=1$ and $f=3$ mean, respectively, the threshold
condition for denoising to be S/N $>$ 1, and 3. Since the real
data of \ion{H}{1} Ly$\alpha$ transmitted flux has high quality,
the power spectrum actually is $f$-independent if $f>1$
\citep{jam05}. On the other hand, there are very rare pixels of
the \ion{He}{2} data with S/N $>$ 3, and therefore, only a very
small number of the modes is available if taking $f>3$. One can
only use $f=1$ to calculate \ion{He}{2} power spectrum of He2347.
The error bars of Figure 7 is estimated by the maximum and minimum
range of bootstrap re-sampling.

As expected, the simulation sample of \ion{H}{1} transmission flux
is in well agreement with observations on all scales less than
$\delta v=224$ km s$^{-1}$. There is no discrepancy on the smallest
scale $\delta v = 28~{\rm km~s^{-1}}$ or length scale
$D = 0.28$ h$^{-1}$ Mpc. This result is the same as Paper I, and
therefore, the power spectrum on small scales is insensitive to the
size of simulation box.

The power spectrum of the \ion{He}{2} Ly$\alpha$ transmitted flux of
HE2347, as shown in the bottom panel of Figure 7, is very
different from simulation samples without adding noise on scales less
than 56 km s$^{-1}$. On those scales, the power spectrum of observed
sample is much higher than the simulation results. However, the power
spectrum of the noisy samples of \ion{He}{2} flux gives a good fitting
to the observed sample. In other words, if we could remove the S/N=3 noise
from the real data, the power spectrum can be well fitted with the
simulation of the LCDM model. Thus, with \ion{He}{2} flux,
we can arrive at the same conclusion as \ion{H}{1}: there is no
evidence for the discrepancy between observation and simulation on
scales from 1792 to 28 km s$^{-1}$, or from 0.28 to 18 h$^{-1}$ Mpc.


\section{Discussions and conclusions}

The \ion{H}{1} and \ion{He}{2} Ly$\alpha$ transmitted flux
fluctuations of QSO absorption spectrum are valuable to detect the
fields of baryon gas and ionizing photon field, and to constrain
models of the UV radiation background. With hydrodynamic
simulation samples of the $\Lambda$CDM model, we made a comparison
between the model-predicted statistical features and real data of
HE2347. It includes the power spectrum, the line width of
absorption features, and the ratio between the \ion{He}{2} and
\ion{H}{1} optical depths. The major conclusions are as follows:

1.) The absorption features of \ion{He}{2} and \ion{H}{1} basically
are turbulent-broadening. It should be emphasized that the observed
evidence for the turbulent-broadening mainly is given by the absorption
lines in voids \citep{zh04}. This is especially supported by the
simulation samples, which shows that the turbulence behavior of the
velocity field of IGM is on high as well as low mass density area
\citep{Kim05}.

2.) The mean of the optical depth ratio $\langle \eta \rangle$ of
    real sample can be well fitted with simulation samples.

3.) The simulation samples give a very well fitting to
the power spectra of both \ion{H}{1} and \ion{He}{2}
transmitted flux. There is no discrepancy between the power
spectra of simulated and observed samples on small scales. For H
sample, the scales are till to 0.28 h$^{-1}$ Mpc. For He, it is
1.1 h$^{-1}$ Mpc. Therefore, there is no evidence that the power
of observed sample is less than the standard $\Lambda$CDM model on
small scales.

4.) The last but not the least is on the $\eta$-scatter.
The large $\eta$ scatter are generally attributed
to the inhomogeneity of the UV photon distributions. We showed,
however, that a significant part of the $\eta$-scatter is from the
fluctuations of the temperature and velocity fields of IGM and
data noise. There seems to be a discrepancy on low-$\eta$ events
comparing our simulation and the real data. The simulation sample
does not contain enough low-$\eta$ events to fit real data.
To solve this problem models of producing extra fluctuations on
Ly$\alpha$ transmitted flux on scales of about 1 h$^{-1}$ Mpc
is needed.

The WIGEON samples are produced with a uniformly distributed UV
radiation background, and therefore, we may consider an
inhomogeneous UV radiation background to be a possible reason of
the lack of low-$\eta$ events. However, the numerical results of
the inhomogeneous UV radiation background shows that the
fluctuations of UV radiation background on small scales will yield
more high-$\eta$ events $(200 < \eta<350)$. Therefore, at least,
according to the current calculation, the model of the UV
background inhomogeneity would not be helpful to solve the problem
of the lack of low-$\eta$ events.

Note that low-$\eta$ events are related to high column density
$N({\rm HI})$. The lack of low-$\eta$ events may be due to that the
WIGEON samples are produced without considering the detailed 
physical processes of ionizing sources, as some mechanical and 
radiation feedback effects would not be negligible in the area of
high column density, which has the high probability containing
collapsed objects.

Any model of producing the scatter of $\eta$ will affect
the transmitted flux of both \ion{H}{1} and \ion{He}{2}.
Therefore, we can use our unified fitting of \ion{H}{1} and \ion{He}{2}
Ly${\alpha}$ transmitted flux to set some constraints on models
for low-$\eta$ events.

First, if the fluctuations are Gaussian, it will play the same
role as Gaussian noise, and increase the power of transmitted flux
on scale $\simeq 1$ Mpc. However, this paper and Paper I show that
the power spectrum of \ion{H}{1} Ly$\alpha$ transmitted flux is in good
consistent with observation on scale 1 h$^{-1}$ Mpc and less.
Therefore, any increase of the power of simulation sample will no
longer be consistent with observed sample. Therefore, there is
small room for models of adding Gaussian fluctuation.

Second, if the fluctuations are non-Gaussian, it should be limited
by the observed non-Gaussian features of the \ion{H}{1} and
\ion{He}{2} flux. The non-Gaussian statistical features
of \ion{H}{1} transmitted flux, such as
high order moments of the fluctuations of the flux, are sensitive
to the addition of non-Gaussian inhomogeneity. It has been shown
in Paper I that the non-Gaussian features of the \ion{H}{1} transmitted
flux of HE2347 can already be well fitted with the WIGEON samples
to eight order and on scale as small as 0.28 h$^{-1}$ Mpc.
Therefore, there is also small room for models of adding
non-Gaussian fluctuations.

\acknowledgments

JRL thanks Francesco Haardt for kindly providing us ionizing
background model, Romeel Dav\'e for making AUTOVP publicly
available and Hy Trac for help from his MACH code. We thank
our referee for helpful suggestions on shaping our paper. 
LLF acknowledges support from the National Science Foundation 
of China (NSFC). This work is partially supported by the
NSF AST-0507340.

\clearpage

\begin{figure}
\figurenum{1}
\epsscale{1.0} \plotone{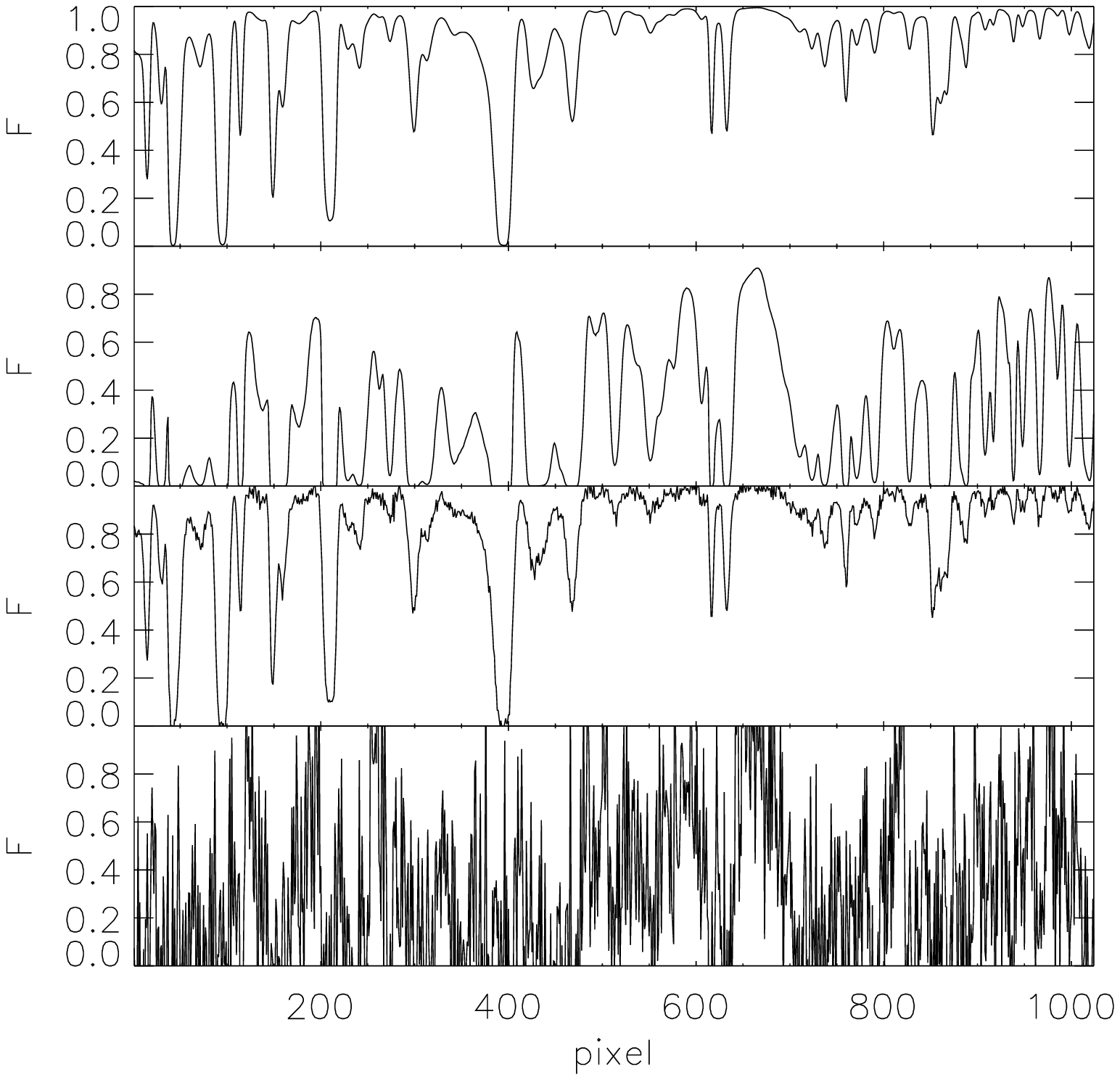}
\caption{Simulated samples of \ion{H}{1} (top) and \ion{He}{2} (second from top)
transmitted flux without adding noise, and the same samples
of \ion{H}{1} (third from top) and \ion{He}{2} (bottom) with adding noise
with S/N=50 for \ion{H}{1} and S/N=3 for \ion{He}{2}.}
\end{figure}

\begin{figure}
\figurenum{2} \epsscale{1.0} \plotone{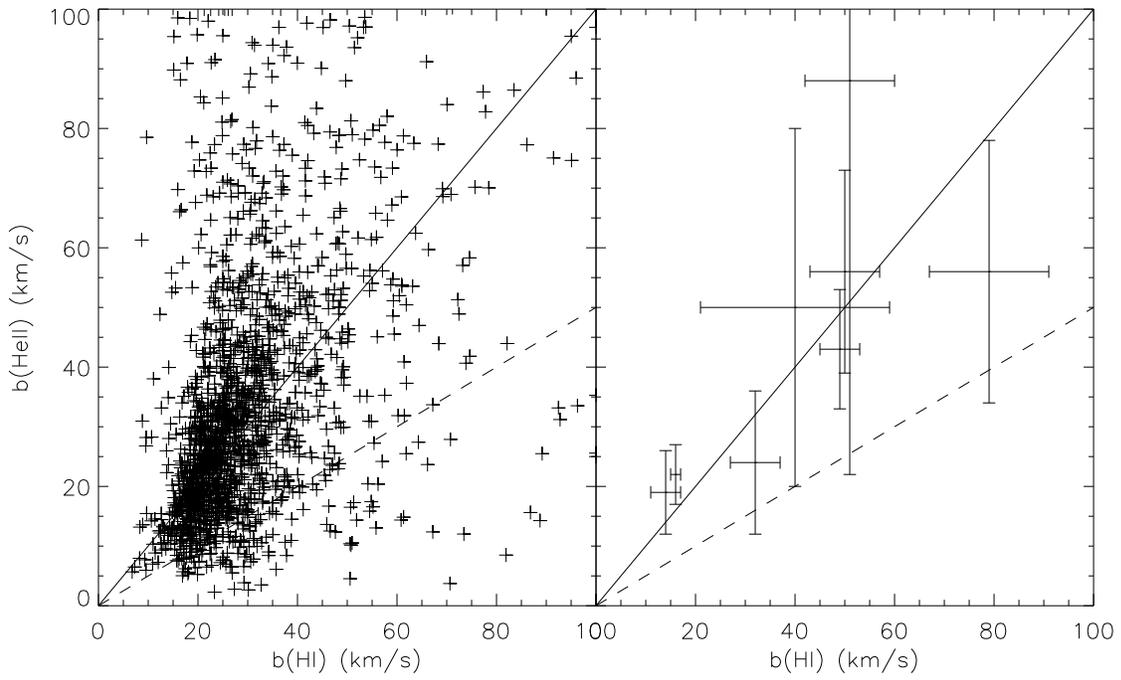}
\caption{Doppler parameter b(\ion{H}{1}) vs. b(\ion{He}{2}) of
simulation samples (left) and real data of HE2347 (right). The
diagonal line is expected for turbulent broadening and the dashed
line is for b(\ion{H}{1}) = 2b(\ion{He}{2}).}
\end{figure}

\begin{figure}
\figurenum{3}
\epsscale{1.0} \plotone{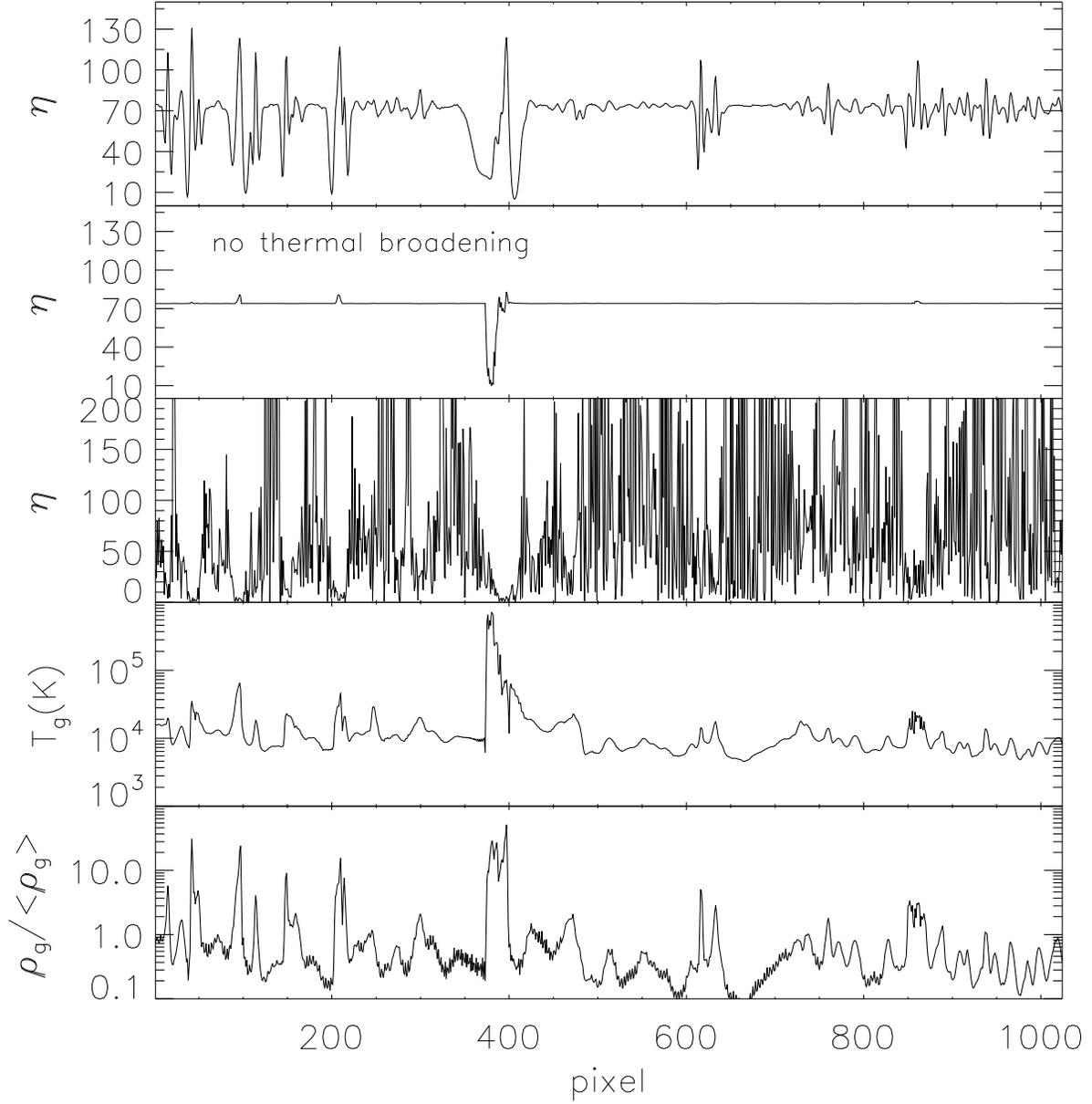}
\caption{1-D distribution of $\eta$ with (top) and without (next top)
thermal broadening.The third panel shows $\eta$ with adding
S/N= 50 noise to \ion{H}{1} and S/N=3 noise to \ion{He}{2} flux.
The temperature and baryon matter density are also shown.}
\end{figure}

\begin{figure}
\figurenum{4}
\epsscale{1.0} \plotone{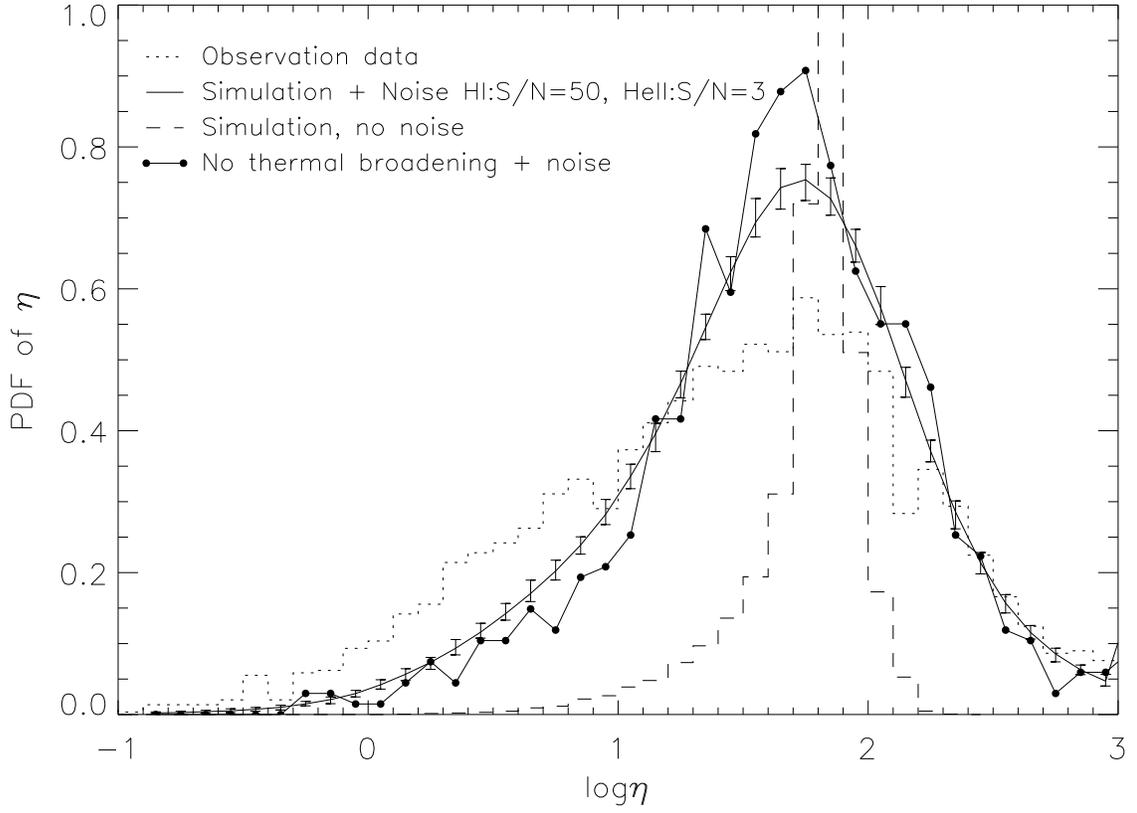}
\caption{PDF of $\eta$. 1. Real data (dot line); 2. Simulation
samples with adding S/N= 50 noise to \ion{H}{1} and S/N=3 noise
to \ion{He}{2} transmitted flux (solid line), errorbars being the maximum
and minimum over 100 independent noise realizations; 3. Simulation
samples without noise (dashed line); 4. Simulation samples 
without thermal broadening, but adding noise.}
\end{figure}

\begin{figure}
\figurenum{5}
\epsscale{0.6} \plotone{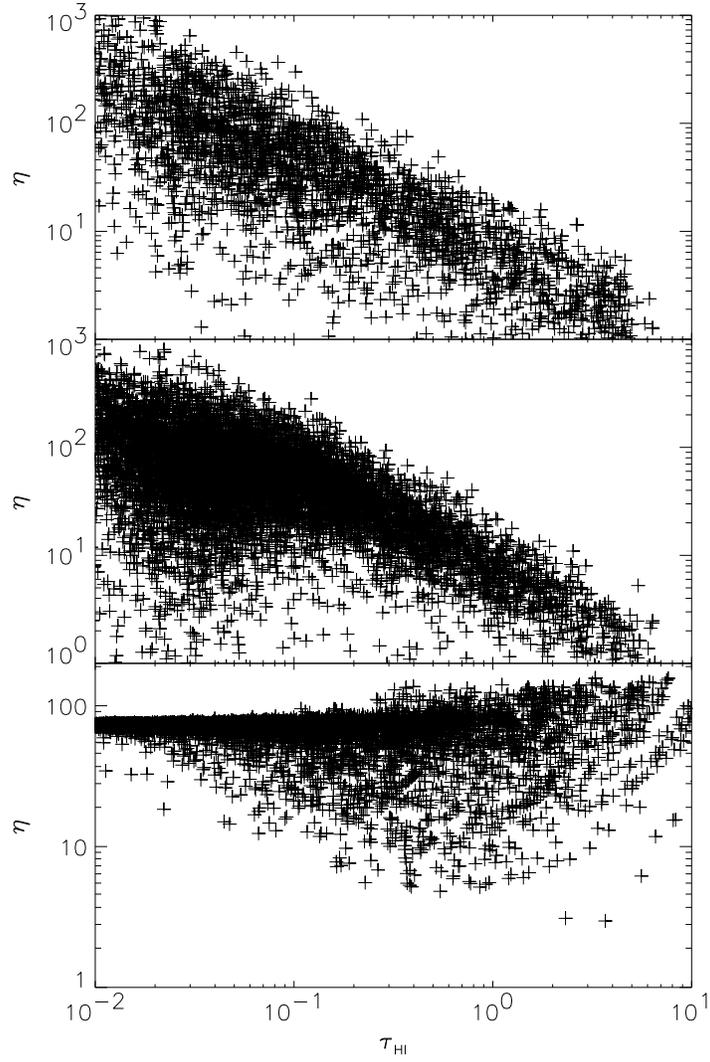}
\caption{$\eta$ vs. optical depth $\tau_{\rm HI}$ for real data (top);
simulation samples with noise (middle) and without noise(bottom).}
\end{figure}

\begin{figure}
\figurenum{6} \epsscale{0.8} \plotone{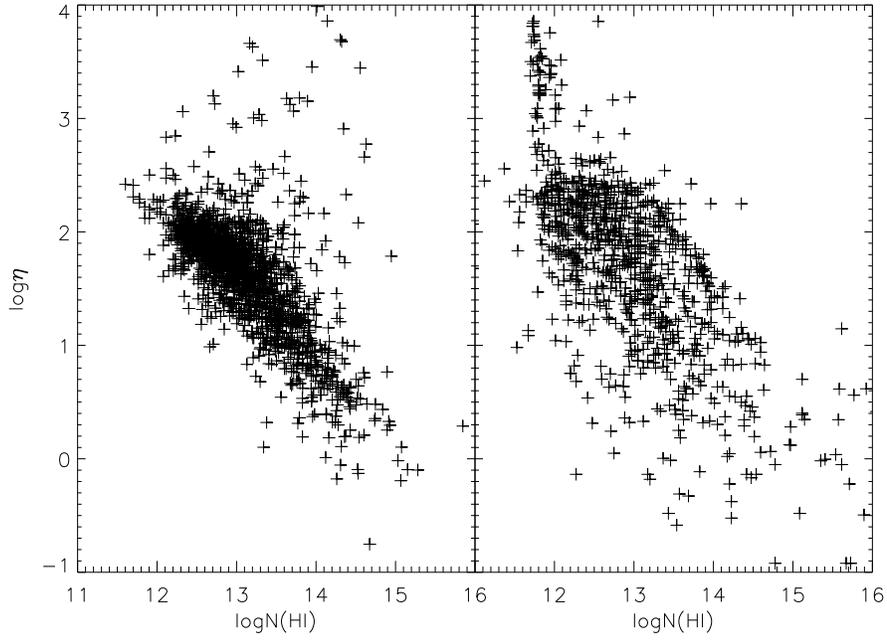}
\caption{ $\eta$ vs. column density $N({\rm HI})$  for simulation
data (left) and real data of HE2347 (right). }
\end{figure}

\begin{figure}
\figurenum{7} \epsscale{0.8} \plotone{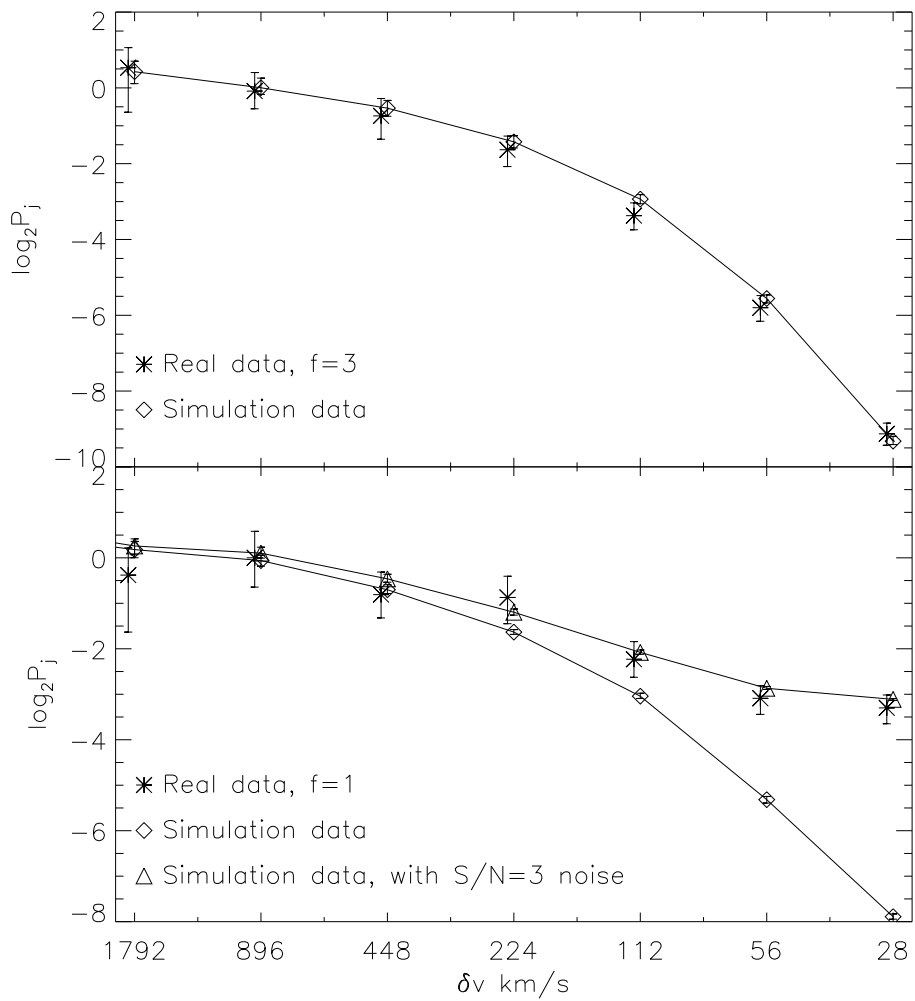} \caption{Power
spectrum of 1.) \ion{H}{1} transmitted flux of HE2347 with the conditional
counting parameter $f=3$ for real data ($\ast$) and simulation
sample ($\Diamond$) (top); 2.) \ion{He}{2} transmitted flux of HE2347
with the conditional counting parameter $f=1$ for real data
($\ast$), simulation sample with adding noise of S/N=3
($\bigtriangleup$) and without noise adding ($\Diamond$)
(bottom).}
\end{figure}

\end{document}